\documentstyle[12pt]{article}

\begin{document}
\begin{center}
{\large GAUSS-BONNET TYPE IDENTITY IN WEYL-CARTAN SPACE}\\
\vskip 0.4cm
O. V. BABOUROVA and B. N. FROLOV \\
\vskip 0.4cm
{\it Department of Mathematics, Moscow State Pedagogical University,\\
     Krasnoprudnaya 14, Moscow 107140, Russia;\\
     E-mail: baburova.physics@mpgu.msk.su,\enspace frolovbn.physics@mpgu.msk.su}
\vskip 1cm
\end{center}

\begin{abstract}
{\small
\par
        The Gauss-Bonnet type identity is derived in a Weyl-Cartan space
on the basis of the variational method.
}
\end{abstract}

\vskip 0.2cm
\section{Introduction}
\markright{Gauss-Bonnet type identity in Weyl-Cartan space}
      In the modern gravitational theory  quadratic Lagrangians are used
widly that is stimulated by the gauge treatment of gravitation and the
renormalization problems in quantum gravity.$^{1}$ In this connection the
Gauss-Bonnet type identity becomes the object of a considerable amount of
attention. The generalization of the Gauss-Bonnet formula to a 4-dimensional
Riemann space $V_{4}$ was performed by Bach $^{2}$ and Lanczos $^{3}$
and on the basis of the variational method by Ray.$^{4}$ The Bach-Lanczos
identity in Riemann spaces imlpies the one-loop renormalizability of pure
gravitation.$^{5}$ The generalization of the Bach-Lanczos identity to a
Riemann-Cartan  space $U_{4}$ was performed in. $^{6-8}$ We shall
obtain the Gauss-Bonnet type identity in a Weyl-Cartan space $Y_{4}$ that can
be essential for the dilatonic gravitational theory with quadratic Lagrangians.
On the preliminary version of our results see Ref. 9.

\section{Preliminaries to the variational procedure}
\markright{Gauss-Bonnet type identity in Weyl-Cartan space}
        We shall consider a Weyl-Cartan space $Y_{4}$ that is a
connected 4-di\-men\-si\-o\-nal oriented differentiable manifold ${\cal M}$ equipped
with a linear connection $\Gamma$ and a metric $g$ (with the Lorenzian
signature) which obey the constraints,
\begin{equation}
Q_{\lambda}\!^{\alpha\beta} = \frac{1}{4}Q_{\lambda}g^{\alpha\beta}\; ,
\quad Q_{\lambda}\!^{\alpha\beta}:= \nabla_{\lambda} g^{\alpha\beta}\; ,
\quad Q_{\lambda}:= Q_{\lambda}\!^{\alpha\beta}g_{\alpha\beta}\;.
\label{eq:1a}
\end{equation}
The tensor $Q_{\lambda}\!^{\alpha\beta}$ is a nonmetricity tensor.$^{10,11}$
\par
 We shall use a holonomic local vector frame $\vec{e}_{\mu}=\vec{\partial}
_{\mu}$ ($\mu=1,2,3,4$) with $\Gamma_{\sigma\rho }\!^{\lambda}$ as a
connection coefficients. A space $Y_{4}$ contains a nonvanishing torsion
tensor, $T_{\sigma\rho}\!^{\lambda}:= 2\Gamma_{[\sigma\rho ]}\!^{\lambda}$,
in general. A curvature tensor of $Y_{4}$ and its various contractions read,
$R_{\mu}\!^{\nu} = R_{\sigma\mu}\!^{\nu\sigma}$, $\tilde{R}_{\mu}\!^{\nu} =
R_{\mu\sigma}\!^{\sigma\nu}$, $R = R_{\sigma}\!^{\sigma}$ and
\begin{equation}
R_{\alpha\beta\sigma}\!^{\lambda} = 2\partial_{[\alpha}\Gamma_{\beta]\sigma}
\!^{\lambda} + 2\Gamma_{[\alpha\vert\rho\vert}\!^{\lambda}\Gamma_
{\beta]\sigma}\!^{\rho}\; . \label{eq:2}
\end{equation}
\par
Let us consider the Lagrangian density,
\begin{equation}
{\cal L}_{0} = \sqrt{-g}L_{0}\; , \quad L_{0} = R^{2} - (R_{\alpha\beta} +
\tilde{R}_{\alpha\beta})(R^{\beta\alpha} + \tilde{R}^{\beta\alpha}) +
R_{\alpha\beta\mu\nu}R^{\mu\nu\alpha\beta}\; .   \label{eq:1}
\end{equation}
The variation of (\ref{eq:1}) with respect to the metric
$g^{\sigma\rho}$ and the connection $\Gamma_{\lambda \nu}\!^{\sigma }$ reads,
\begin{equation}
\delta{\cal L}_{0} = -\frac{1}{2}\sqrt{-g}H_{\sigma\rho}\delta g^{\sigma
\rho} - \sqrt{-g}H_{\sigma}\!^{\nu\lambda}\delta\Gamma_{\lambda\nu}\!^
{\sigma} + \mbox{total divergence} \; , \label{eq:3}
\end{equation}
where
\begin{eqnarray}
\sqrt{-g}H_{\sigma\rho}: = -2\left[ \frac{\delta{\cal L}_{0}}
{\delta g^{\sigma\rho}}\right]_{\Gamma=const}\;,\quad  \sqrt{-g}H_{\sigma}
\!^{\nu\lambda}:= -\left[\frac{\delta{\cal L}_{0}}{\delta\Gamma_{\lambda\nu}
\!^{\sigma}}\right]_{g^{\sigma\rho}=const}\;, \label{eq:4} \\
H_{\sigma\rho} = g_{\sigma\rho}L_{0} - 4R_{\alpha\beta(\sigma}\!^{\tau}
R_{\rho )\tau}\!^{\alpha\beta} - 4R_{\tau (\sigma\rho )}\!^{\kappa}(R_{\kappa}
\!^{\tau} + \tilde{R}_{\kappa}\!^{\tau}) \nonumber \\
- 4 R_{\tau (\sigma}(R_{\rho )}\!^{\tau} + \tilde{R}_{\rho )}\!^{\tau}) -
4RR_{(\sigma\rho )}\; , \label{eq:5}\\
\sqrt{-g}H_{\sigma}\!^{\nu\lambda}= 4\nabla_{\mu} \{ \sqrt{-g} [R_{\sigma}\!^
{\nu [\lambda\mu ]} + (R_{\sigma }\!^{[\lambda} +\tilde{R}_{\sigma }\!^
{[\lambda})g^{\mu ]\nu} \nonumber \\
- (R^{\nu [\lambda} + \tilde{R}^{\nu [\lambda})\delta_{\sigma}\!
^{\mu]} - R\delta_{\sigma}^{[\lambda}g^{\mu ]\nu} ]\} \nonumber\\
+ 2\sqrt{-g} [ R_{\sigma}\!^{\nu\alpha\beta} + (R_{\sigma}\!^{\alpha} +
\tilde{R}_{\sigma}\!^{\alpha})g^{\beta\nu} - (R^{\nu\alpha} + \tilde{R}^
{\nu\alpha})\delta_{\sigma}^{\beta} + Rg^{\nu\alpha}\delta_{\sigma}^{\beta}
] M_{\alpha\beta}\!^{\lambda}\; . \label {eq:6}
\end{eqnarray}
Here the modified torsion tensor is introduced, $M_{\sigma\rho}\!^{\lambda}:=
T_{\sigma\rho}\!^{\lambda} + 2\delta_{[\sigma}^{\lambda}T_{\rho]}$, where
$T_{\rho} := T_{\rho\tau}\!^{\tau}$ is the torsion trace.

\section{The Gauss-Bonnet type Teorem \newline in Weyl-Cartan space}
\markright{Gauss-Bonnet type identity in Weyl-Cartan space}
      The following Gauss-Bonnet type Teorem generalized to $Y_{4}$ is valid.
\par
{\em Theorem}: The integral quantity,
\begin{equation}
\int_{{\cal M}} \, \sqrt{-g} [R^{2} - (R_{\alpha\beta} + \tilde{R}_{\alpha
\beta})(R^{\beta\alpha} + \tilde{R}^{\beta\alpha}) + R_{\alpha\beta\mu\nu}
R^{\mu\nu\alpha\beta}] \, d^{4}x \; ,   \label{eq:18}
\end{equation}
over the oriented 4-dimensional manifold ${\cal M}$ without boundary equipped
with the Weyl-Cartan differential-geometric structure does not depend on the
choice of a metric and a connection of the manifold and is a topological
invariant.
\par
        {\em Proof}. The main idea of the proof consists in the demonstration
that the variation of the integrand of (\ref{eq:18}) with respect to a
metric and a connection in a Weyl-Cartan space  $Y_{4}$ is equal identically
to a total divergence. In  $Y_{4}$ the metric and the connection are not
independent because of the constraints (\ref{eq:1a}). Therefore one can vary
the modified integrand expression,
\begin{equation}
{\cal L} = {\cal L}_{0} + \frac{1}{2} \sqrt{-g}H_{\alpha\beta}\!^{\lambda}
\left (Q_{\lambda}\!^{\alpha\beta} - \frac{1}{4}Q_{\lambda}g^{\alpha\beta}
\right )\; , \label{eq:9}
\end{equation}
which in $Y_{4}$  coinsides  with the integrand of (\ref{eq:18}).
\par
    The variation of (\ref{eq:9}) has the form,
\begin{eqnarray}
&\delta{\cal L} = -\frac{1}{2}\left (\sqrt{-g}H_{\sigma}\!^{\nu} +
\stackrel{*}{\nabla}_{\lambda}\left [\sqrt{-g}(H_{\sigma}\!^{\nu\lambda}
- \frac{1}{4}\delta_{\sigma}^{\nu} H_{\tau}\!^{\tau\lambda})\right ]
\right ) g_{\nu\rho}\delta g^{\sigma\rho}\nonumber \\
&+\frac{1}{2}\sqrt{-g} \left ( H_{\sigma}\!^{\nu\lambda}(Q_{\lambda\nu\rho}-
\frac{1}{4}Q_{\lambda}g_{\nu\rho}) - \frac{1}{2}g_{\sigma\rho}
H_{\alpha\beta}\!^{\lambda}(Q_{\lambda}\!^{\alpha\beta}-\frac{1}{4}
Q_{\lambda}g^{\alpha\beta}) \right )\delta g^{\sigma\rho} \nonumber \\
& - \sqrt{-g}\left (H_{[\sigma\rho ]}\!^{\lambda} + \frac{1}{4}g_{\sigma\rho}
H_{\tau}\!^{\tau\lambda}\right )g^{\rho\nu}\delta\Gamma_{\lambda\nu}\!^
{\sigma} \nonumber \\
& + \frac{1}{2} \sqrt{-g} \left (Q_{\lambda}\!^{\alpha\beta} -\frac{1}{4}
Q_{\lambda}g^{\alpha\beta}\right )\delta H_{\alpha\beta}\!^{\lambda}
+ \mbox{total divergence} \; , \label{eq:90}
\end{eqnarray}
where $\stackrel{*}{\nabla}_{\lambda} = \nabla_{\lambda} + T_{\lambda}$.
If the constraints (\ref{eq:1a}) are taken into account, then the hypothesis
of the Theorem  is the consequence of the Lemma.
\par
        {\em Lemma}: In a Weyl-Cartan space $Y_{4}$ the following identities
are valid,
\begin{equation}
(a)\,\;\sqrt{-g}H_{\sigma}\!^{\rho} + \stackrel{*}{\nabla}_{\lambda}
(\sqrt{-g}H_{\sigma}\!^{\rho\lambda})= 0\;,\quad (b)\,\;H_{[\sigma\rho ]}
\!^{\lambda}= 0\;,\quad (c)\,\;H_{\tau}\!^{\tau\lambda}=0\;. \label{eq:13}
\end{equation}
\par
        {\em Proof}. The statement (c) follows immediately from (\ref{eq:6}).
Using (\ref{eq:6}) and the Bianchi identities in $Y_{4}$ one gets,
\begin{eqnarray}
H_{[\sigma\nu ]}\!^{\lambda} = [ 8R_{\tau [\sigma}\!^{[\lambda\alpha ]}
\delta_{\nu ]}^{\beta} + 4R_{\tau [\sigma}\!^{\alpha\beta}\delta_{\nu]}^{\lambda}
+ 4(R_{\tau}\!^{\alpha} + \tilde{R}_{\tau}\!^{\alpha})\delta_{[\sigma}^{\beta}
\delta_{\nu ]}^{\lambda}  \nonumber \\
+  2(R_{\tau}\!^{\lambda} + \tilde{R}_{\tau}\!
^{\lambda})\delta_{\sigma}^{\alpha}\delta_{\nu}^{\beta}] T_{\alpha
\beta}\!^{\tau} \nonumber \\
+ [ 2R_{\sigma\nu}\!^{\alpha\beta} + 4(R_{[\sigma}\!
^{\alpha} + \tilde{R}_{[\sigma}\!^{\alpha})\delta_{\nu ]}^{\beta} - 2R\delta
_{\sigma}^{\alpha}\delta_{\nu}^{\beta}] M_{\alpha\beta}\!^{\lambda}\; .
\label{eq:14}
\end{eqnarray}
Let us consider the expression,
\begin{equation}
B_{\sigma\nu}\!^{\lambda} = \frac{1}{2}\eta_{\alpha\beta\kappa\omega}
\eta^{\mu\tau\kappa\epsilon}\eta^{\phi\rho\omega\lambda}
\eta_{\delta\sigma\epsilon\nu}R_{\phi\rho}\!^{\alpha\beta}
T_{\mu\tau}\!^{\delta}\; , \label{eq:15}
\end{equation}
where $\eta_{\alpha\beta\gamma\delta}=\sqrt{-g}e_{[\alpha\beta\gamma\delta]}$
is the Levi-Civita tensor ($e_{[1234]} = -1$), and calculate this expression
in two ways: the first way consists in combining the first factor with the
second one and the third factor with the forth one, while the second way
consists in combining the first factor with the third one and the second
factor with the forth one. By equating the two results one gets that the
statement (b) of the Lemma is valid.
\par
        With the help of the Bianchi identities one can find,
\begin{eqnarray}
\sqrt{-g}H_{\sigma}\!^{\nu} + \stackrel{*}{\nabla}_{\lambda}(\sqrt{-g}H_
{\sigma}\!^{\nu\lambda}) =  \sqrt{-g} [\delta_{\sigma}^{\nu}L_{0} +
4R_{\sigma\tau}\!^{\alpha\beta}R_{\alpha\beta}\!^{[\tau\nu ]} \nonumber \\
 + 4R_{\sigma}\!^{\kappa [\tau\nu ]}(R_{\tau\kappa} + \tilde{R}_{\tau
\kappa}) + 2(R_{\sigma}\!^{\tau} +\tilde{R}_{\sigma}\!^{\tau})
(R_{\tau}\!^{\nu} + \tilde{R}_{\tau}\!^{\nu}) -2R(R_{\sigma}\!^{\nu} +
\tilde{R}_{\sigma}\!^{\nu})]\; . \label{eq:16}
\end{eqnarray}
Let us consider the expression,
\begin{equation}
B_{\sigma\rho} = \frac{1}{4}\eta_{\delta\tau\phi}\!^{\psi}\eta_{\mu\nu}\!^
{\phi\omega}\eta_{\alpha\beta\psi\rho}\eta_{\kappa\epsilon\omega\sigma}
R^{\alpha\beta\delta\tau}R^{\mu\nu\kappa\epsilon}\; , \label{eq:17}
\end{equation}
and calculate it in two ways, as before. After equating the two results one
can be convinced that the statement (a) of the Lemma is valid. The proof
of the Lemma is finished.
\par
Using the Lemma one can see that in $Y_{4}$ the variation of (\ref{eq:90})
and therefore the variation of (\ref{eq:18}) is equal to a total divergence.
Q.E.D.

\vskip 0.6cm
\par {\large\bf References}
\vskip 0.4cm
\begin{description}
\item
1. K.S. Stelle, {\em Phys. Rev.} {\bf D16}, 953 (1977).
\item
2. R. Bach, {\em Math. Z.} {\bf 9}, 110 (1921).
\item
3. C. Lanczos, {\em Ann. Math.(N.Y.)} {\bf 39}, 842 (1938).
\item
4. J.R. Ray, {\em J. Math. Phys.} {\bf 19}, 100 (1978).
\item
5. G. 't Hooft and M. Veltman, {\em Ann. Inst. H. Poincar\'{e}} {\bf 20}, 69
(1974).
\item
6. V.N. Tunjak, {\em Izvestija vyssh. uch. zaved. (Fizika)} N9, 74 (1979) [in
Russ.].
\item
7. H.T. Nieh, {\em J. Math. Phys.} {\bf 21}, 1439 (1980).
\item
8. K. Hayashi and T. Shirafuji, {\em Prog. Theor. Phys.} {\bf 65}, 525 (1981).
\item
9. O.V. Babourova, S.B. Levina and B.N. Frolov, in {\em Teoreticheskie i
eksperimental\'{}nye problemy gravitacii} (Abstracts of Contr. Pap. 9 Russ.
Grav. Conf., Novgorod) (Moscow, 1996), Part 1, p. 43 (in Russian).
\item
10. J.A. Schouten, {\em Ricci-Calculus} (Springer, Berlin, 1954).
\item
11. F.W. Hehl, J.D. McCrea, E.W. Mielke and Yu. Ne'eman, {\em Phys. Reports}
{\bf 258}, 1 (1995).
\end{description}
\end{document}